\documentclass[preprint, review, 12pt, 3p]{elsarticle}

\usepackage[table,xcdraw]{xcolor}
\usepackage{todonotes}
\usepackage{comment}
\usepackage{lipsum}
\usepackage{amsmath}
\usepackage[hidelinks]{hyperref}
\usepackage{graphicx}
\usepackage{subfig}
\usepackage{multirow}
\usepackage{lipsum}
\usepackage{multirow}
\usepackage{dirtytalk}
\usepackage{caption}
\usepackage{adjustbox}
\usepackage{tabularx}
\usepackage{floatrow} 
\usepackage[vlines]{tabularht}
\usepackage{pbox}
\usepackage{booktabs}

\usepackage{float}
\floatstyle{plaintop}
\restylefloat{table}
\usepackage{rotating}
\usepackage[linesnumbered,ruled,vlined]{algorithm2e}
\SetKwInput{KwInput}{Input}                
\SetKwInput{KwOutput}{Output}              

\usepackage{hyperref}
\usepackage{academicons}
\usepackage{xcolor}

\usepackage{hyperref}
\usepackage{xcolor}
\definecolor{lime}{HTML}{A6CE39}
\DeclareRobustCommand{\orcidicon}{%
    \begin{tikzpicture}
    \draw[lime, fill=lime] (0,0) 
    circle [radius=0.16] 
    node[white] {{\fontfamily{qag}\selectfont \tiny ID}};
    \draw[white, fill=white] (-0.0625,0.095) 
    circle [radius=0.007];
    \end{tikzpicture}
    \hspace{-2mm}
}

\foreach \x in {A, ..., Z}{%
    \expandafter\xdef\csname orcid\x\endcsname{\noexpand\href{https://orcid.org/\csname orcidauthor\x\endcsname}{\noexpand\orcidicon}}
}
\newcommand{\orcid}[1]{\href{https://orcid.org/#1}{\textcolor[HTML]{A6CE39}{\orcidicon}}}

\newsavebox\MBox

\journal{arXiv}

\begin{document}

\begin{frontmatter}

\title{EEG based stress analysis using rhythm specific spectral feature for video game play}

\author[label1]{Shidhartho Roy \orcid{0000-0001-8448-0790}} 
\ead{swapno15roy@gmail.com}

\author[label1]{Monira Islam \corref{cor1}}
\ead{monira@eee.kuet.ac.bd}

\author[label1]{Md. Salah Uddin Yusuf}
\ead{suyusuf@eee.kuet.ac.bd}

\author[label1]{Nushrat Jahan}
\ead{nushrat739@gmail.com }

\address[label1]{Department of Electrical and Electronic Engineering, Khulna University of Engineering \& Technology, Khulna-9203, Bangladesh}

\cortext[cor1]{Corresponding author}

\begin{abstract}

\subsection*{Background and Objective}
For the emerging significance of mental stress, various research directives have been established over time to better understand the causes of stress and how to deal with it.
In recent years, the rise of video gameplay is unprecedented, further triggered by the lockdown imposed due to the COVID-19 pandemic. Several researchers and organizations have contributed to the practical analysis of the impacts of such extended periods of gameplay, which lacks coordinated studies to underline the outcomes and reflect those in future game designing and public awareness about video gameplay. 
Investigations have mostly focused on the “gameplay stress” based on physical syndromes only. Some studies have analyzed the effects of video gameplay with EEG, MRI, etc., without concentrating on the relaxation procedure after video gameplay. 

\subsection*{Methods} 
This paper presents an end-to-end stress analysis for video gaming stimuli using EEG. The PSD value of the Alpha and Beta bands is computed to calculate the Beta-to-Alpha ratio (BAR). In this article, BAR is used to denote mental stress. Subjects are chosen based on various factors such as gender, gameplay experience, age, and BMI. EEG is recorded using Scan SynAmps2 Express equipment. There are three types of video gameplay: strategic, puzzle, and combinational. Relaxation is accomplished in this study by the use of music of various pitches. Two types of regression analysis is done to mathematically model stress and relaxation curve. Brain topography is rendered to indicate the stressed and relaxed region of the brain.

\subsection*{Results} 
In the relaxed state, the subjects have BAR $0.701$, which is considered the baseline value. Non-gamer subjects have an average BAR of $2.403$ for 1 hour of strategic video gameplay, whereas gamers have $2.218$ BAR concurrently. After $12$ minutes of listening to low-pitch music, gamers achieved $0.709$ BAR, which is nearly the baseline value. In comparison to Quartic regression, the $4PL$ symmetrical sigmoid function performs regression analysis with fewer parameters and computational power.

\subsection*{Conclusion}
Non-gamers experience more stress than gamers, whereas strategic game creates more stress on human brain. During gameplay, the beta band in the frontal region is mostly activated. For relaxation, low pitch music is the most useful media. Residual stress is evident in the frontal lobe when the subjects experienced high pitch
music. Quartic regression performs regression analysis follow the relaxation curve more accurately compared to 4PL symmetrical sigmoid function.\\

\end{abstract}

\begin{keyword}
Beta-Alpha ratio \sep EEG \sep Stress - relaxation modeling \sep  Topography \sep Video gameplay.
\end{keyword}

\end{frontmatter}

\section{Introduction}
\label{sec:introduction}


In the modern society, stress is a significant matter of concern. According to WHO, loss of productivity and mental health disorders have globally cost a significant capital \cite{mozos2017stress}. When the human body is exposed to negative stimuli, it responds with stress. Acute stress is the result of short-term stressful events, which are typically indicated by transient physiological changes. Acute stress lasting for a long time might turn into episodic stress. Chronic stress, anxiety, and clinical depression can result from prolonged exposure to stress factors or traumatic situations \cite{rai2012psychological}. This paper mainly centers around acute stress. To study acute stress, video gameplay has been employed as the stimulus.


Video games (VG) are becoming increasingly popular among young people in current culture, and VG are growing in popularity not just as a research tool but also as an area of study. In addition, a plenty of research is focused on the neurological and behavioral consequences of VG because of the expanding popularity and universal applications, offering a comprehensive brain correlation in recent decades. Moreover, the rising diversity of digital technology, such as smartphones and tablets, has brought entertainment software or VG into the mainstream. Thus, a significant portion of society (more than 30\% in tablets and 70\% in smartphones) is exposed to these technologies and may now, in some way, be termed as casual gamers \cite{choi2020commercial}.

Because of its high temporal resolution, non-invasiveness, and low cost, electroencephalography (EEG) is frequently utilized in research, including neural engineering, neuroscience, and biomedical engineering (e.g., brain-computer interfaces, BCI \cite{he2018brain}, sleep analysis \cite{motamedi2014signal}, stress analysis \cite{yusuf2019stress}, and seizure detection \cite{chen2014automatic}). EEG has provided insight into the brain's functioning in many cognitive processes and is regarded as the building block of functional signaling in the brain. Oscillations of various frequencies are linked to various cognitive functions. Alpha oscillations, for example, are linked to relaxation, whereas beta rhythms are linked to tension.

When it comes to relaxation, listening to music is unquestionably a top contender. Music has a powerful impact on both the mind and the body. Faster music can help us to concentrate and feel more alert. Upbeat music can boost one's mood and make one feel more optimistic about life. A slower tempo can relax the mind and muscles, making a person feel soothed while letting go of the day's stress. This study uses three different types of music (low, medium, and high pitch) to study relaxation.

\begin{figure}[!ht]
 \centering
\includegraphics[width=1\linewidth]{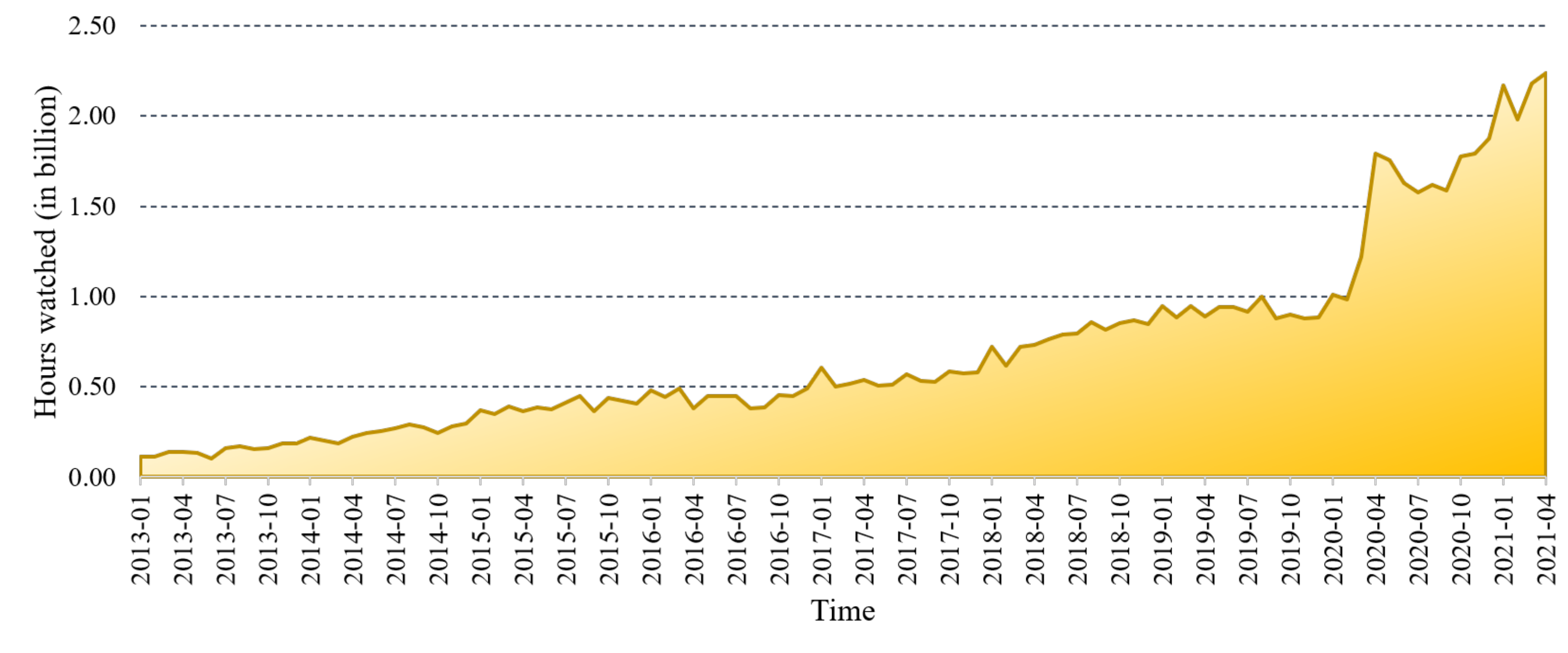}
  \caption{The surge of gameplay hours watched since 2013. At the beginning of the pandemic, the gameplay hours watched have almost doubled \cite{TWITCH_2021}.}
  \label{fig:game_covid}
\end{figure}

In the mainstream media, both good and bad health claims associated with VG are common. These comments are primarily unconfirmed and sensational, based on 'expert' opinions, but without proof. 
Therefore, it is becoming increasingly important to know the actual impacts of long-term exposure to video gaming in a deeper technical context. 
The effect can be positive in the form of cognitive, motivational, emotional, and social benefits \cite{zayeni2020therapeutic} or can be negative in the form of exposure to graphic violence, addiction, contribution to obesity, and cardiometabolism \cite{kracht2020video}.
For a long time, video gaming has been on the increase. However, the growth in gaming was unprecedented during the COVID pandemic. According to a gamesindustry.biz study, VG sales have surged by $63$ percent in $50$ major areas since pre-pandemic time \cite{zhu2021psychology}. The home quarantine compelled everyone to explore the world of gaming, resulting in an $80\%$ increase in the number of new games downloaded.
Under the pandemic, the social infrastructure for individuals, particularly those in quarantine, has evolved into various platforms \cite{marston2020role}.
The rise of gameplay in recent years is evident from Fig. \ref{fig:game_covid}. In early 2020, when lockdown started, the duration of video gameplay in hours has almost doubled from before.

Some researchers have previously focused on the influence of VG, depending on participants' exposure to that particular game or studies, on the content features of VG \cite{kuhn2014playing}. Some research specializes in the genres of VG with prominent graphic violence (action, war, shooting, killing, etc.) On this basis, a reasonable quantity of scientific documentation on the research of violence in VG and violence acclimatization is used \cite{martinez2021exploring}.
For decades, social, political, and scientific considerations have been given to the link between exposure to media violence and violent behavior. Extensive empirical data suggest that exposure to media violence impacts aggressive behavior, abuse, violent cognition, physiologic excitement, and teenage animosity \cite{wang2021differences}. 
Another study has shown that playing some VG, particularly "action" games, often improves the cognitive function \cite{martinez2021exploring}.
It has also given thought on how the business may be better regulated to promote games for cognitive advancement. Considerable research has been conducted on the stressful effects of video gaming on the human brain \cite{johnstone2020development}. Using EEG as a tool for study, Soysal et al. distinguished the function of the brain for different games and the stress level caused by them. It can differentiate between a relaxed and a stressed individual \cite{soysal2020quantifying}.
Saputra et al. researched a crucial aspect of video gaming and found that, in addition to the entertainment value of video gaming, players get stressed when they experience frustration as a result of virtual failure \cite{saputra2017stress}.
  
 \begin{figure}[!ht]
 \centering
\includegraphics[width=0.5\linewidth]{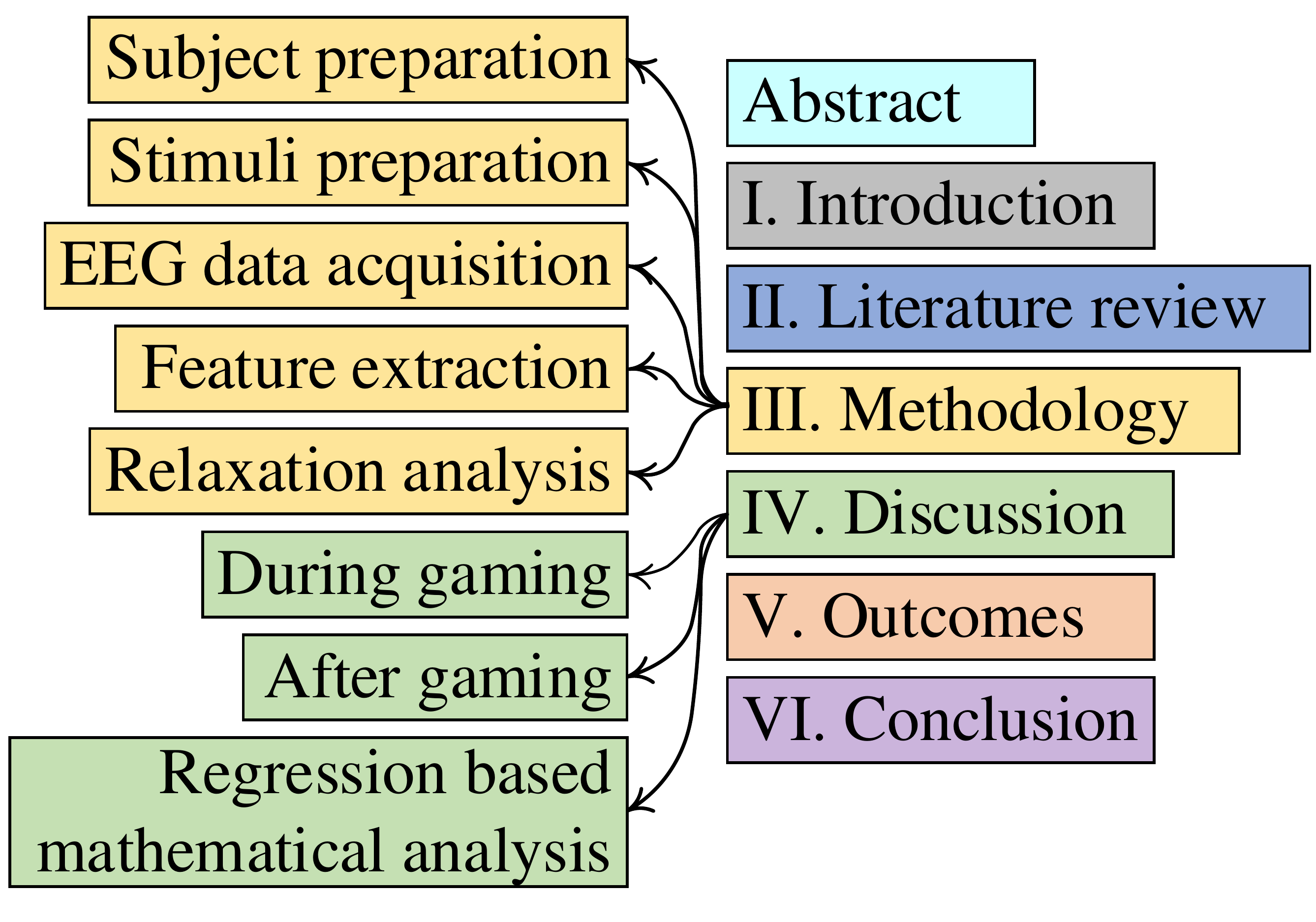}
  \caption{Paper organization structure.}
  \label{fig:paper_structure}
\end{figure} 
 

\begin{table*}[!ht]
\tiny
\caption{EEG based stress analysis in the recent literature.}
\label{tab:lit_review}
\begin{tabular}{|m{1.45cm}|m{2.45cm}|m{2.45cm}|m{2.45cm}|m{5.45cm}|m{0.45cm}|}
\hline
\textbf{Bio-medical signal type} & \textbf{Notable Methodology}                                                    & \textbf{Stimuli}                                  & \textbf{Subject}                                        & \textbf{Contribution}                                                                                                                                            & \textbf{Ref}              \\ \hline
EEG                              & Stress-related hormone labeling, ML-based classification                        & construction work                                 & Workers  at Construction Sites                          & Developed a procedure to recognize construction worker’s stress by applying supervised learning algorithms         & \cite{jebelli2018eeg}            \\ \hline
EEG                              & Multiclass SVM with error correcting output code (ECOC) & arithmetic task                                   & 20 to 24 years old male right-handed adults             & The classification of stress with a precision of 94.79\% shows the right prefrontal cortex dominance over mental stress.                            & \cite{al2018towards}             \\ \hline
EEG                              & Frequency analysis from average power                                           & Multiplayer Online Battle Arena Video Game & 18-25 years old 10 participants                         & First time experienced participant have significant difference for alpha/beta ratio than experienced gamer.                                                      & \cite{saputra2017stress}         \\ \hline
EEG                              & Unity3D game engine, EMOTIV kit                                                 & Video game                                        & ADHD patients                                           & EEG-based game that targets ADHD symptomatic individuals to augment their focus.                                                                                 & \cite{alchalcabi2017more}        \\ \hline
EEG and fNIRS                    & SVM classifier, decision fusion of fNIRS and EEG signals                        & Montreal Imaging Stress Task (MIST)               & 22 male, right-handed adults (age 22 ± 2 years) & suggesting fNIRS+EEG as an efficient tool of stress analysis                                                                                                   & \cite{al2017stress}              \\ \hline
EEG, PPG and GSR                 & Keystroke Analysis                                                              & Video game                                        & 22 subjects (9 female) having mean age 29.3 years       & Analysis of keystroke data gathered in the tests to suggest a new method of stress measurement                                                          & \cite{das2017classification}     \\ \hline
EEG                              & power spectral density (PSD)                                                    & to prepare a talk on an unknown topic.            & 28 participants                                         & Finding an appropriate EEG recording phase for classifying mental stress into many groups.                                                          & \cite{arsalan2019classification} \\ \hline
EEG                              & PSD, Wavelet transformation, topographic analysis                               & Music of different genres.                        & 48 individuals were selected from 11 to 50 years of age & Short length of music offers alleviation from tension, however the stress is growing when individuals have been exposed to music for a long period.           & \cite{roy2019age}                \\ \hline
EEG                              & KNN, LDA, MLP                                                                   & TempleRun video game                              & 20 healthy subjects                                     & A game player's competence level is classified on a consumer gaming device                                                                        & \cite{anwar2018game}             \\ \hline
EEG                              & CBA Ratio and the Statistical Analysis                                          & Music                                             & nine participants                                       & In a stressful scenario, a drop of brain alpha wave frequencies and a concurrent increase in the participant blood pressure and pulse is shown & \cite{paszkiel2020impact}        \\ \hline
EEG                              & Genetic Algorithm-Based Feature Selection                                       & 40 music videos                                   & Thirty-two healthy participants                         & classification performance is best when the proposed GA-based method was used.                                                                                   & \cite{shon2018emotional}         \\ \hline
EEG, ECG                         & Analog amplification in the front-end                                              & Colors and their names                            & Seven healthy young male subjects                       & Long-term daily monitoring is needed to evaluate chronic stress                                                                                 & \cite{ahn2019novel}              \\ \hline
EEG                              & PSD, beta/alpha ratio                                                           & Language based sustained mental task              & 52 subjects all aged from 20 to 23                      & Mother language creates the least stress.                                                                                                                        & \cite{yusuf2019stress}           \\ \hline
EEG                              & Wavelet, PSD                                                                    & Language based music                              & 30 subjects all aged from 20 to 23                      & Bass frequency reveals more relaxation.                                                                                                                          & \cite{roy2020frequency}          \\ \hline
\end{tabular}
\end{table*}

Despite these researches, relatively few works are based on relaxing the stress caused by video gaming.
The motivation behind this study was to resemble real-world scenarios as EEG stimuli and analyze video gameplay with EEG. For this purpose, we conducted an open survey on the nature of video gaming on various people. Among 214 participants, $41.35\%$ indicated they do not do anything after gaming, while $38.67\%$ stated they like some music after a prolonged gaming period. Because these two populations outnumber others, we pooled them in our study. To select the video gaming period, we asked our participants how long they play at a time. On average, they reported that they play 55.6 minutes of VG. Some gamers also documented that they play 3 to 4 hours at a stretch, and this has been considered as outlier in this study. The relaxation period of 12 minutes is also based on the survey. Subjects were asked to choose how long they want to spend time after video gameplay before starting another cognitive intensive work. The average response time was 11.08 minutes.

Furthermore, a few studies have been discovered to develop a mathematical model based on stress development from gaming and the relaxation via music of various pitches. 
Our goal in this study is mental state analysis during gameplay of various game genres, brain mapping of these states, mental state analysis after playtime, relaxing via multiple forms of music, and corresponding regression analysis. This is, to the best of our knowledge, the first attempt to aggregate all of them  \footnote{After acceptance, code and dataset will be made open access for further research in the link: \url{https://github.com/royShidhartho/Stress-analysis-video-gameplay}}.
The rest of the paper is organized as shown in Fig. \ref{fig:paper_structure}.

\section{Literature Review}
\label{sec:lit_review}
From the very beginning of the research to examine the response of the brain due to video gameplay, the effects of video gameplay using EEG has been emphasized. Previously some of the studies have been focused on other techniques rather than EEG. But EEG had good accuracy in examining brain activity for video game playing. From 2017 to 2021, many developments took place in this era, such as using fMRI, MRI, PPG, GSR, fNIRS, ECG, etc. But accuracy was not good compared to EEG. Though ECG is the technique that is used the second-highest for this purpose, it is actually a low-quality method. Furthermore, it is difficult to eliminate noise with  conventional filtering methods for fNIRS. From the table \ref{tab:lit_review} below, it is clear that the EEG technique is best suited for examining the stress analysis in different time duration.

\begin{figure}[!ht]
 \centering
\includegraphics[width=1\textwidth]{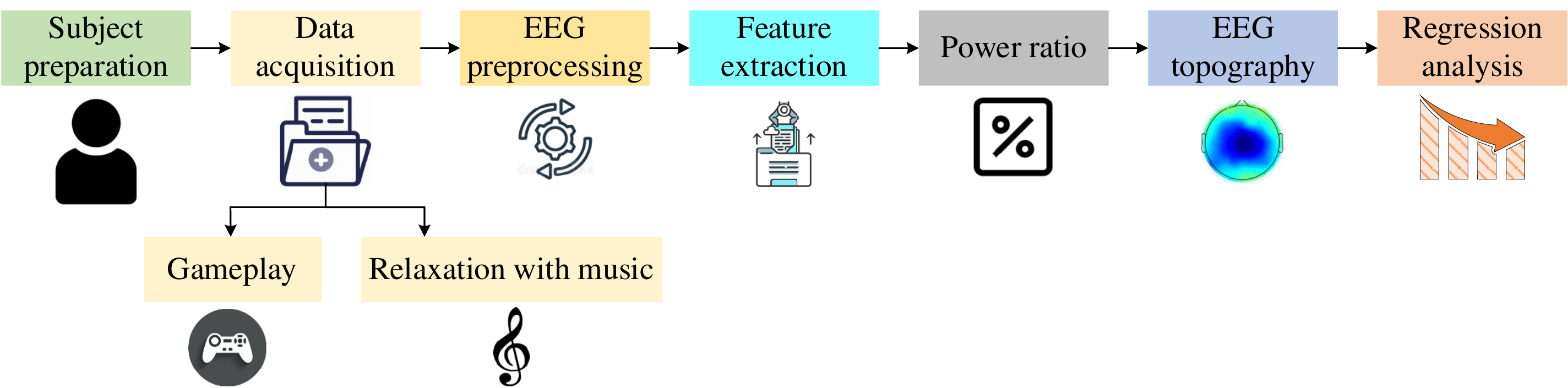}
  \caption{Block diagram of this experiment.}
  \label{fig:block}
\end{figure}

Another assessment observed from the literature review is that people aged 19 to 24 years are mostly impacted by video gaming. This circumstance has both positive and negative impacts on society. Though video gameplay can improve problem-solving skills, it is time-consuming. We had also observed that old adults, as well as patients, are involved in playing gameplay. The peak age ranged from 19 to 24 years and with increasing age, the observation and the interest in video gaming becomes lower consequently.

\section{Methodology}
\label{sec:Methodology}
To achieve the objective of this experiment, several methods are used. Subjects were first prepared after settling on a data acquisition protocol using an EEG recording device EEG signal and pre-processing by both onboard and digital devices. The power ratio was computed after extracting features from the clean EEG signal. This experiment made use of a variety of games and music. Following the experiment, brain mapping was created using the EEG topography approach, and mathematical modeling was completed using regression analysis. Fig. \ref{fig:block} illustrates the block diagram of this experiment.

\subsection{Subject Preparation}
\begin{figure}[!ht]
 \centering
\includegraphics[width=0.5\linewidth]{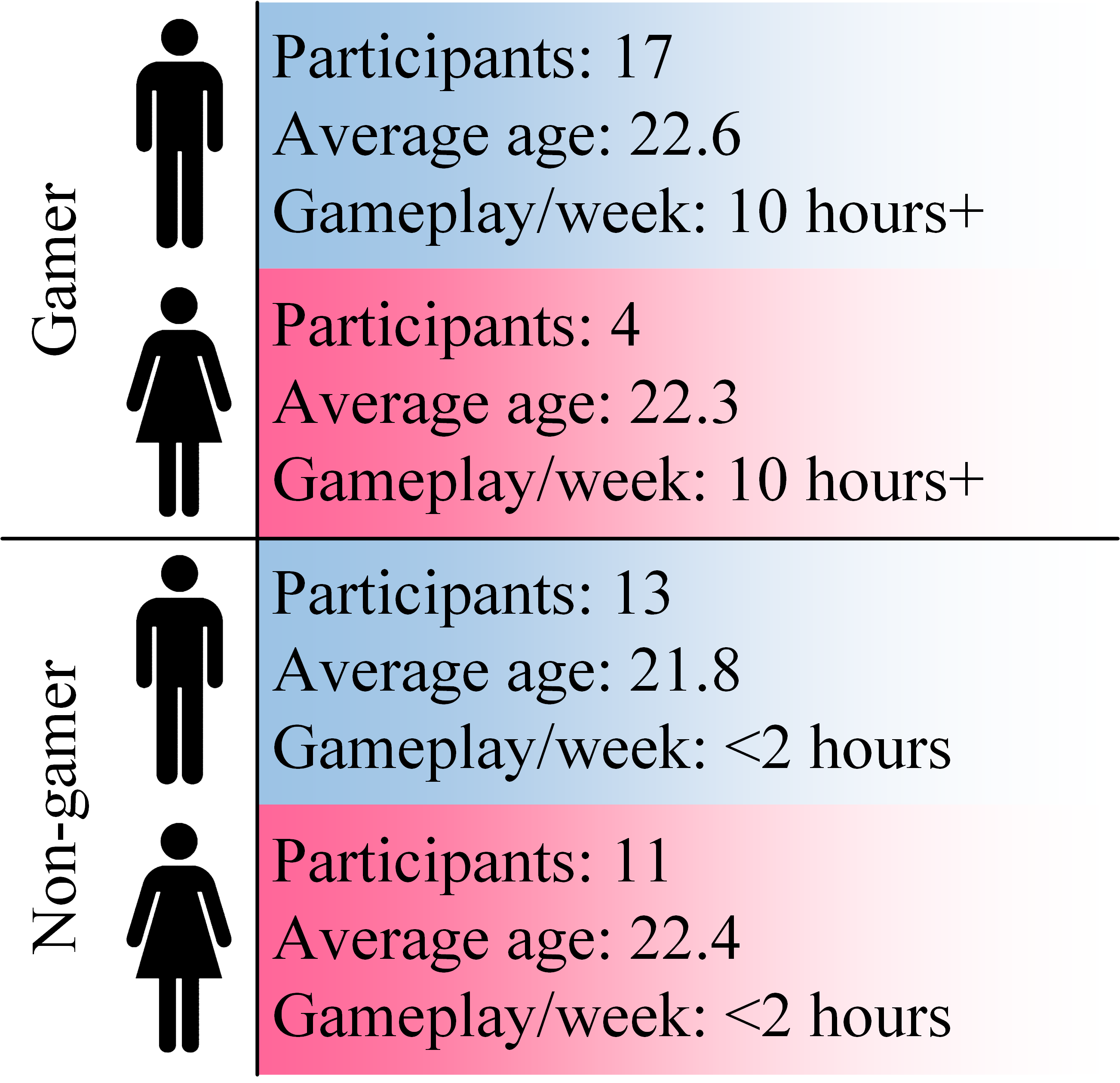}
  \caption{Subject distribution of this study. Blue shades denote the male subjects, and pink shade embodies the female subjects.}
  \label{fig:subjectsv2}
\end{figure}

Four focus groups are organized for this study ($FG_1$ – $FG_4$) and from four to $17$ each ($N = 45$). Participants were selected using an open survey procedure based on their experience in gaming and their health condition. All subjects are between the ages of 20 and 24 years (22.29 on average) and university students. These are organized into four focus groups: two groups are called 'Gamers' who on average play games 10 hours a week or more. The other two groups play games for less than 2 hour per week or do not play at all.
Both men and women participated in the gamer and non-gamer groups. The proportion of women in non-gamers is 45.83\%, compared to 19.04\% in gamers. There is also a lack of sleep disorders or other serious disorders in the selected subjects. Furthermore, the subjects were unaware of the study's goals. In figure \ref{fig:subjectsv2}, the subject distribution is depicted. Another criteria for selecting the subjects was to select them from a wide BMI spectrum. Eight of the subjects in this study are obese (BMI 30.0 and above), 18 of them are overweight (BMI 25.0-29.9), 12 of them are normal (BMI 18.5-24.9), and the rest of them, 7, are underweight (below BMI 18.5).

\subsection{Stimuli preparation}

VG is used as stimuli to create stress in this study. Based on game design, all VG were divided into three groups. On one end of the scale, puzzle games demand quick thinking, whereas strategy games need quick response and extended pondering.
\begin{enumerate}
    \item \textbf{Puzzle:} Video puzzle games aim to solve riddles. Logic, pattern recognition, sequence resolution, and word completion are only a few of the problem-solving capabilities that these puzzles might test for. The player may have endless time or endeavors to solve a puzzle or a time restriction, or it may make it more challenging to do fundamental problems by having to do it in real-time.
    
    \textit{Gameplay format:} These games usually include rules in which players use a grid, network, or other interactive areas to control game pieces. In order to earn a triumph, the players must unravel clues so that they may then move to the next level. Usually, completing every riddle will lead to a more complicated task.
    
    \item \textbf{Strategy:} A video strategy game promotes strategic thinking and preparation to win. It focuses on strategic, tactical, and sometimes logistical questions. In this area, there are additional action games. Many games also feature economic concerns and exploration. It is classified into four subtypes depending on whether the game is in turn or in real-time, whether strategic or tactical—for example, Rome: Total War, Dota 2, etc.
    
    \textit{Gameplay format:} In order to reduce opposing troops, a player must perform a sequence of operations against one or more opponents. Victory is achieved through a greater plan with less luck. The user receives a divine vision of the game world, with most strategy VG offering them indirect authority over the game units. Consequently, most strategy games involve a mixture of tactical and strategic considerations and contain various fighting components.

    \item \textbf{Combinational: } It accumulates both the puzzle and strategy type games. Such as Grand Theft Auto, San andreas etc.
    
    \textit{Gameplay format:} A combination game implies a timed sequence of plays, each of which leaves the opponent incapable or near impossible to block or otherwise prevent the next succession.
    
\end{enumerate}

\begin{figure}[!ht]
 \centering
\includegraphics[width=0.5\linewidth]{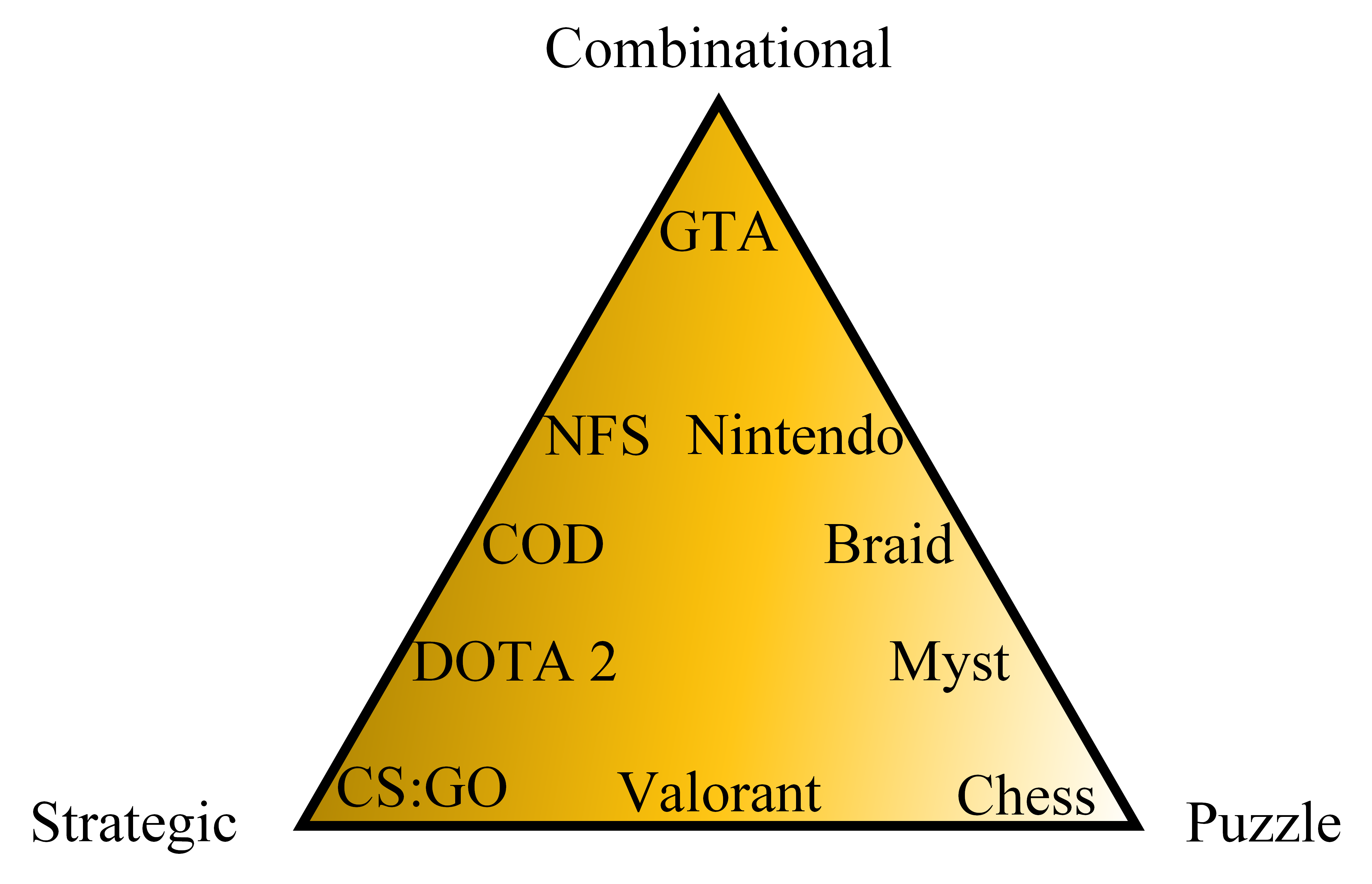}
  \caption{Game classification considered for this study.}
  \label{fig:game_classification}
\end{figure}

\subsection{EEG data Acquisition}

\begin{figure}[!ht]
 \centering
\includegraphics[width=0.6\linewidth]{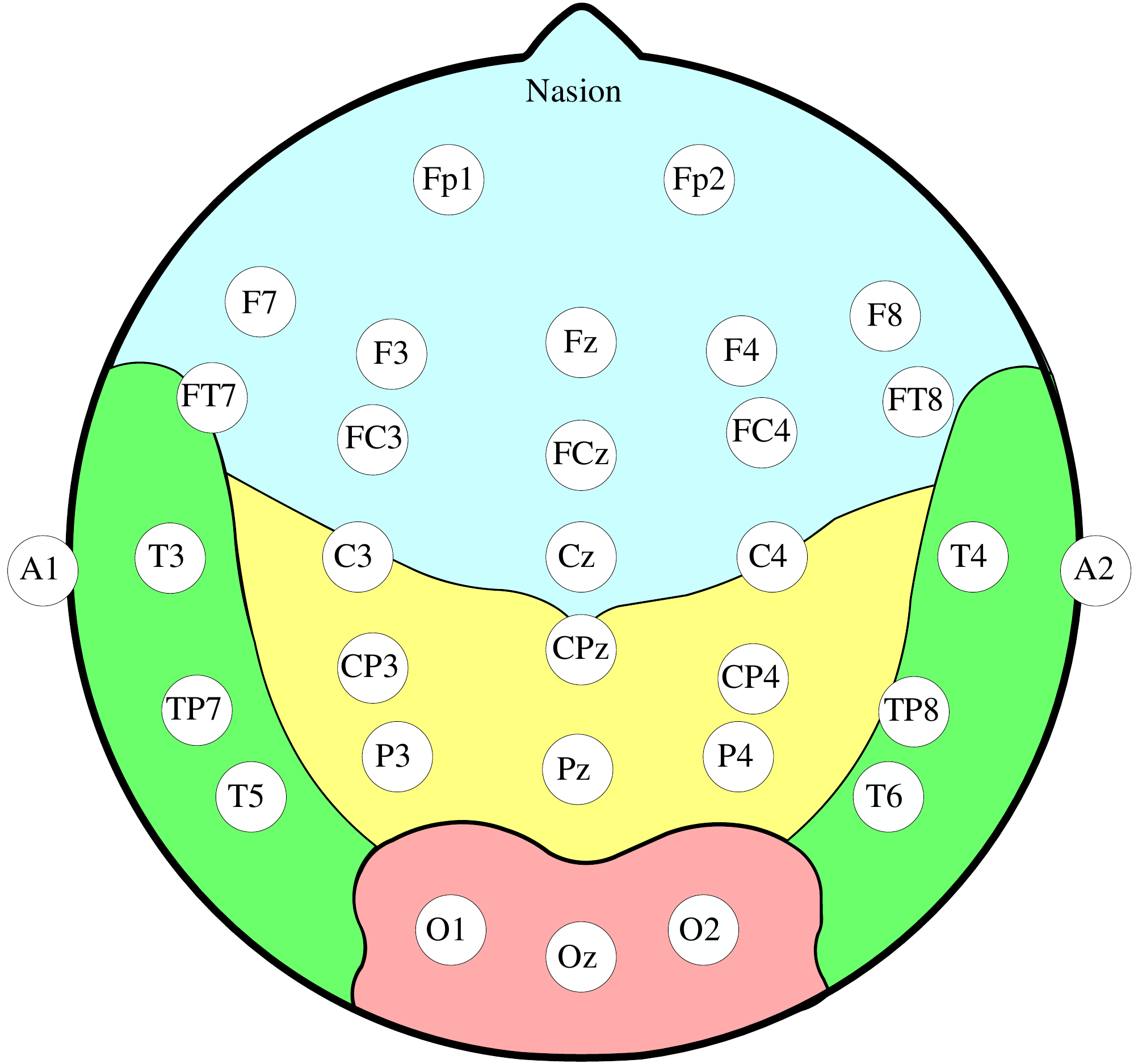}
  \caption{10-20 system of electrode placement. Blue electrode denotes the electrode positions used in this experiment.}
  \label{fig:10-20}
\end{figure}

The Scan SynAmps2 Express equipment (Compumedics Ltd., VIC, Australia) was used to acquire EEG signals. A wired EEG cap containing 32 Ag/AgCl electrodes, comprising 30 EEG electrodes and two reference electrodes, was utilized to acquire EEG data (opposite lateral mastoids). An adapted international 10–20 system was used to put the EEG electrodes. All electrodes and the skin had a contact impedance of less than $5\ k\Omega$ . The EEG recordings were amplified and digitized at 500 Hz using the Scan SynAmps2 Express device (resolution: 16 bits). Scan 4.5 from Neuroscan is the data collecting tool. On the computer, the raw data was stored as .cnt files.
Figure \ref{fig:10-20} manifests the the experiment electrode positions.

\begin{figure}[!ht]
 \centering
\includegraphics[width=0.4\linewidth]{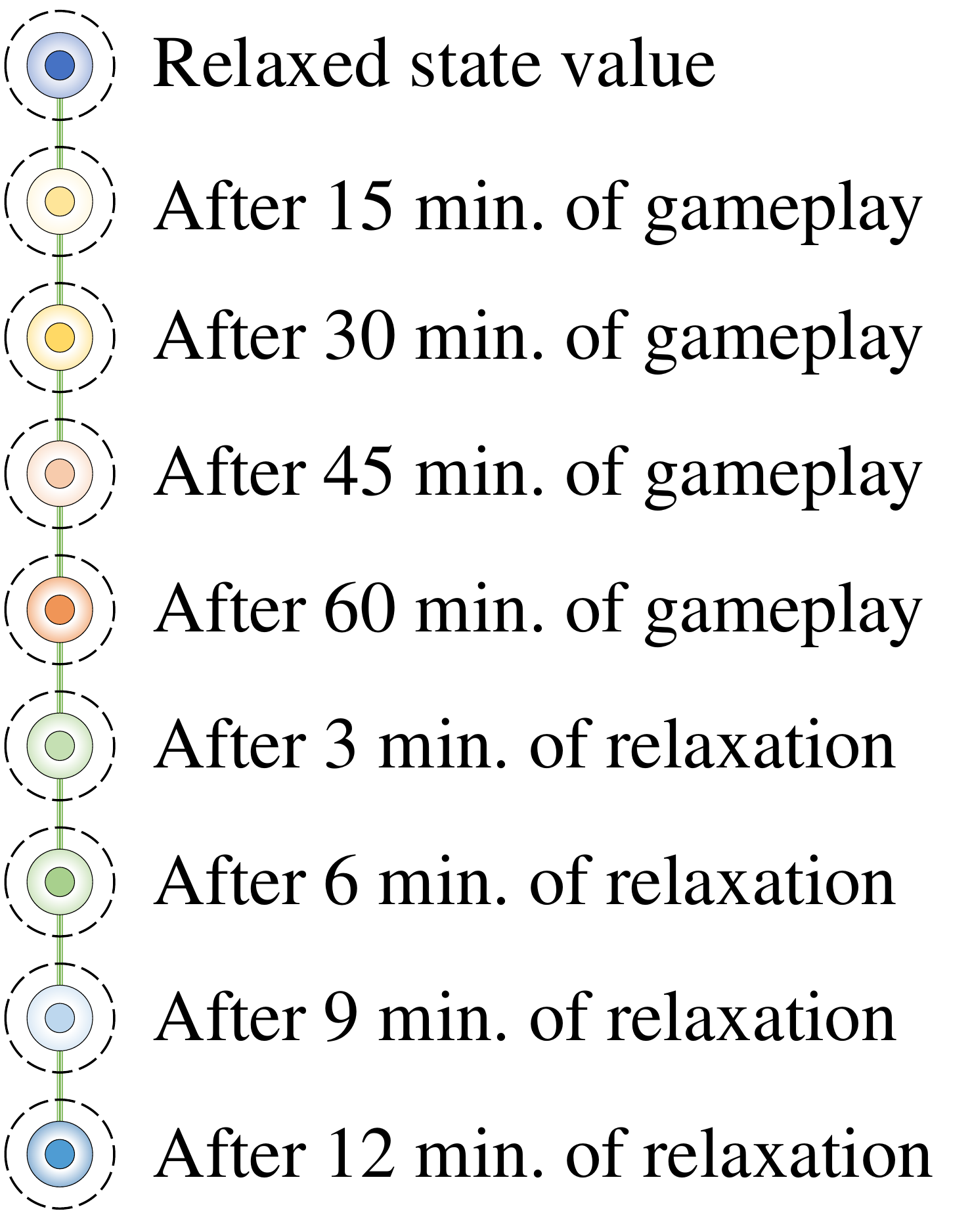}
  \caption{Data acquisition protocol. The orange color palette denotes the during gameplay phase, and the green color palette represents after gameplay phase.}
  \label{fig:acquisition}
\end{figure}

The acquisition of data is divided into two streams – one during video gameplay and one following gameplay. In the gameplay stage, the relaxed state value was captured initially after 10 minutes of relaxation in experimental set-up data.
The data was collected during the games in an interval of 15 minutes. Data were recorded after the gameplay within a period of 3 minutes. The workflow of data gathering for the experiment is displayed in the Fig. \ref{fig:acquisition} .

\subsection{Feature extraction}

The FFT is a sophisticated reversible mapping technique used to transfer a signal from a time domain to a frequency domain. In the periodogram calculated by the FFT technique, the PSD properties of artifact-free EEG data may be evaluated. However, there is elevated volatility and low statistical precision in the periodogram. Welch developed a modified periodogram averaging approach, resulting in a reduction in PSD variance \cite{welch1967use}.

This process involves splitting time series data into segments, overlapping them, and then conducting FFT, calculating and averaging the amplitude of each segment. In this study, 72 minutes EEG signals were selected as samples, and the window length of each 10 seconds was divided into four equal segments. The time sequence was separated into $50\%$ for each sample of 10 seconds, and four segments were overlapped. For each segment, FFT was used to calculate and average the capacity of the overlapping FFT coefficients in all segments. This calculation is done for all frequency ranges until, ultimately, the Alpha and Beta PSD ($\mu V^2 / Hz$) values have been generated. The PSD is manifested according to Welch in the following equation:

\begin{equation}
\label{eq:welch1}
    P_d(f) = \frac{1}{MU} \sum\limits_{n=0}^{M-1} \left| x_d  (n) w(n) e^{-j 2 \pi f n}\right|^2, \ 
\end{equation}

for $d\ =$ $1,\ 2,\ 3, ...,\ L$ signal intervals, $x_d  (n)$ is the sequence, where M is the interval length. $w(n)$ represents the window data, whereas $U$ is the normalization factor for window power. $U$ is defined as
\cite{subha2010eeg},

\begin{equation}
\label{eq:welch2}
    U = \frac{1}{M} \sum\limits_{n=0}^{M-1} \left| w(n) \right|^2. \ 
\end{equation}

The Welch spectrum of power can be shown as the mean over these modified periodograms:

\begin{equation}
\label{eq:welch3}
    P_{Welch}(f) = \frac{1}{L} \sum\limits_{i=0}^{L-1} P_d(f), \ 
\end{equation}

where $ P_{Welch}$ is each interval's periodogram of the EEG signal.

\textbf{Rhythm power ratio}
The power ratio was calculated by dividing the absolute power of one frequency band by the absolute power of another frequency band, using the averaged PSD of each frequency band. Delta activity revealed the deep resting condition of a person. It was not presumed to display substantial activity in the face of stress, which is why this rhythm has not been studied.
By dividing the PSD of beta by the PSD of alpha, the beta/alpha rhythm power ratio was determined. After that, it was evaluated with the ratio value of the difference between the baseline and the stress session.
The processes can also be used to determine the Theta/Beta ratio and differentiate between relaxed and stressed states.The path of data collecting, signal pre-processing, feature extraction, and power ratio computation is depicted in Figure \ref{fig:block}.

\subsubsection{EEG topography}
EEG topography is a sort of neuroimage including the geometrical design of uniformly spaced sites of multiple EEG electrodes on the head. A certain computer software maps the activity on the color screen or printer in many tones of colors by sorting the number of actions (for example, blue might represent low EEG amplitude, while yellow and red might represent larger amplitudes). From EEG topography, we are exploring the residual stress on the human brain. Residual stress is termed as the remaining stress after the designed relaxation period for this study.

In mathematical interpolation techniques (calculation of intermediate values on the values of their neighbors), the spatial locations between electrodes are established to provide a smooth color gradation. This technique delivers the position of rhythm, amplitude changes regarding the skull surface, and a considerably more precise and realistic image.

Comparing topographies for pattern similarity in various settings can demonstrate various thinking processes and so directly test cognitive theories. Topographic analysis of variance (TANOVA) is a statistical test that examines similarity topographies. In this topographic similarity measure, called "angle measure," the topographical pattern similarity is measured by a high dimensional angle between two topographies.

Multivariate topography patterns across all sensors are displayed as $\vec{A}$ and $\vec{B}$ vectors with a size equal to that of the two situations. The following equation (Eq. \ref{eq:topo}) calculates an angle cosine of $\theta$ to quantify the topographical similitude between the two scenarios.


\begin{equation}
\label{eq:topo}
    \cos \theta = \frac{\vec{A} . \vec{B}}{\left| \vec{A} \right|.\left| \vec{B} \right|} ,
\end{equation} 

The cosine value is an index that shows the spatial correlation between two circumstances in which the \textit{+1} value indicates the same patterns as the \textit{-1} value. In addition, because the magnitude of the response in the two circumstances normalizes this index, it benefits that the magnitude of the responses does not impact it.

\subsubsection{Regression analysis and metrics}
Two types of regression analysis is employed in this paper.
\begin{itemize}
  \item \textbf{4PL Symmetrical sigmoid function}
This results in a more advanced model that is better suited to many biological processes. The four-parameter logistic regression is known as this model (4PL). It is very beneficial for dose-response or receptor binding tests or comparative tests. It contains four parameters to be evaluated to 'fit the curve,' as the name suggests. The model is designed for data making an S-shaped curve. The model's equation is:  

\begin{equation}
\label{eq:regression}
    y = d + \frac{a-d}{1+(\frac{x}{c})^b} ,
\end{equation}  

where ,
a = the least value that can be achieved (i.e. what happens at relaxed state.), 
d = the highest value that can be achieved (i.e. what happens after 1 hour of gameplay),
c = inflection point (i.e. the point on the S shaped curve halfway between a and d), and
b = Hill’s slope of the curve (the steepness of the curve at point c).

Fig. \ref{fig:sigmoid} denotes the structure of the 4PL Symmetrical sigmoid function.
  
  \begin{figure}
 \centering
\includegraphics[width=0.5\linewidth]{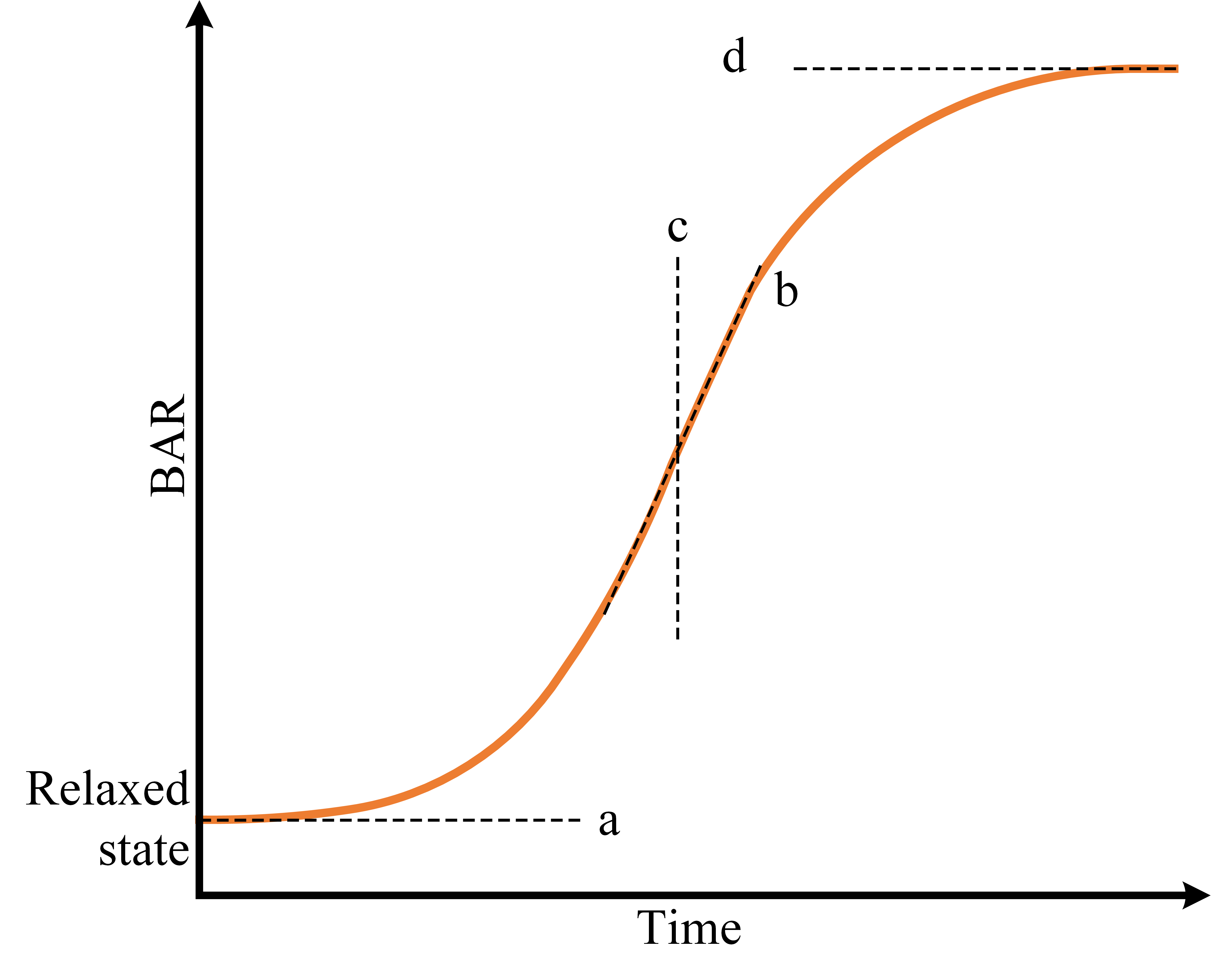}
  \caption{Symmetrical sigmoid function.}
  \label{fig:sigmoid}
\end{figure}

  \item \textbf{Quartic Regression}
The quartic regression adapts to a data set with a quartic function (a polynomial function with a degree of 4). The form of quartic function is: 
  
\begin{equation}
\label{eq:regression1}
    y = a + bx + cx^2+ dx^3+ ex^4+ fx^5 
\end{equation}    
The quartic function takes a range of forms with various inflection points (when a function changes form) and zero to many roots (places where the graph crosses the axis). If the f value exceeds 0, three distinct fundamental forms can be present.
  
\end{itemize}
 
\textbf{Metrics}
Two types of regression metrics are employed here to fit the stress and relaxation curve for this experiment.
\begin{itemize}
  \item \textbf{Coefficient of determination}

A statistical measure of how close the data are to the correct regression line is R-squared. It is also known as the determination coefficient or the multiple regression coefficients. A linear model may explain a certain proportion of the variation in a response variable:

\begin{equation}
\label{eq:regression3}
    R^2 = 1 -  \frac{sum\ of\ the\ squares\ of\ residuals}{total\ sum\ of\ squares}
\end{equation} 
A high $R^2$ number, in general, suggests that the model is a good fit for the data, though fit interpretations vary depending on the context of study. For example, a $R^2$ of 0.44 shows that 44 percent of the variation in the outcome can be explained simply by predicting the outcome using the covariates in the model. In certain subjects, such as the social sciences, that proportion may be difficult to forecast; in others, such as the physical sciences, one would expect $R^2$ to be considerably closer to 100\%. 

  \item \textbf{Akaike information criterion (AIC)}

The AIC is a prediction error estimator and a comparable quality for a certain number of data models. AIC determines the quality of each model concerning each of the other models based on the collection of models for the data \cite{mcelreath2018statistical}. AIC offers a way to pick the model. The information theory is the basis of AIC. When used to describe the process that produced the data, a statistic model is virtually never accurate; hence, information is lost through the usage of the model to describe the process. The AIC calculates the amount of information lost by a particular model: the less information lost, the greater the model's quality \cite{taddy2019business}.

\begin{equation}
\label{eq:regression4}
 AIC = 2k-2ln(\hat{L}),
\end{equation}

where $\hat{L}$ is the maximum value of the likelihood function for the model and k is the number of estimated parameters in the model.
AIC considers the trade-off between model goodness of fit and model simplicity when assessing the amount of information lost by a model. To put it another way, AIC considers both the risks of overfitting and underfitting.

\end{itemize}

\subsection{Relaxation analysis}

Music is used as a relaxation media in this study. Pitch has been used to categorize music in this research.
Pitch is the characteristic that allows people to perceive sounds as "upper" or "lower" in the sense of musical melodies. It is a perceptual feature of sounds that allows them to be arranged on a frequency-related scale[1]. Only sounds with a precise and stable frequency that can be differentiated from noise can calculate pitch. Pitch is a significant auditory attribute of musical tones, along with duration, loudness, and timbre.

Pitch can be described as a frequency; however, it is a subjective psychoacoustical aspect of sound rather than a completely objective physical characteristic. Pitch has long been a critical topic in psychoacoustics, with theories of sound encoding, processing, and perception in the auditory system being created and tested.
The frequency of vibration of the sound waves that create them determines the pitch of sounds. A high pitch is defined as a frequency of 880 hertz [Hz; cycles per second], while a low pitch is defined as a frequency of 55 hertz [Hz; cycles per second]. The music is divided into the following categories:
\begin{itemize}
  \item Low pitch music: below 200 Hz,
  \item Medium pitch music: 200 to 600 Hz, and
  \item High pitch music: above 600 Hz
\end{itemize}
To bring down the beta band activity in previous levels, subjects are divided into 4 groups.
  \begin{itemize}
  \item \textbf{Group 1:} No music was introduced to bring down the brain activity to normal.
  \item \textbf{Group 2:} Low pitch music was introduced to bring down the brain activity to normal.
  \item \textbf{Group 3:} Medium pitch music was introduced to bring down the brain activity to normal
  \item \textbf{Group 4:} High pitch music was introduced to bring down the brain activity to normal.
\end{itemize}

\section{Experimental result analysis}
The full experimental study is divided into three phases; stress level observation during gaming, relaxation level determination after video gameplay and mathematical modeling of brain stress and relaxation. After discussing all the observations, the summary of the discussion's are pointed out in evaluation sub-section.

\subsection{Stress analysis}
Before going into the gameplay and data acquisition, the relaxed state alpha and beta value are measured. On average, the subjects exhibit BAR as 0.701 (Alpha = 4.329, Beta = 3.034). This value from now on is considered as the baseline value.
When the gameplay starts, subjects have gone through 3 distinct set of VG; puzzle, combinational, and strategic. As the subjects are broadly divided into 2 groups, gamer and non-gamer, each group’s data are calculated in average. For data acquisition, 15 minutes interval is considered.

\begin{table*}[!ht]
\scriptsize
\caption{Change of mental states during gameplay, observed with BAR.}
\label{tab:mental_states}
\begin{tabular}{|m{1.2cm}|m{1.2cm}|m{1.2cm}|m{1.3cm}|m{1.2cm}|m{1.3cm}|m{1.2cm}|m{1.3cm}|m{1.2cm}|m{1.3cm}|}
\hline
\textbf{Game Types}            & \textbf{Gamer type} & \textbf{After 15 min} & \textbf{Increase (\%)} & \textbf{After 30 min} & \textbf{Increase (\%)} & \textbf{After 45 min} & \textbf{Increase (\%)} & \textbf{After 1 Hour} & \textbf{Increase (\%)} \\ \hline
\multirow{2}{*}{Puzzle}        & Gamer               & 0.729                 & 0.0384                 & 0.874                 & 0.1979                 & 1.297                 & 0.4595                 & 1.541                 & 0.5451                 \\ \cline{2-10} 
                               & Non-Gamer           & 0.862                 & 0.1868                 & 0.984                 & 0.2876                 & 1.542                 & 0.5454                 & 1.897                 & 0.6305                 \\ \hline
\multirow{2}{1.45cm}{Combi- national} & Gamer               & 0.814                 & 0.1388                 & 0.917                 & 0.2356                 & 1.341                 & 0.4773                 & 1.797                 & 0.6099                 \\ \cline{2-10} 
                               & Non-Gamer           & 1.084                 & 0.3533                 & 1.295                 & 0.4587                 & 1.742                 & 0.5976                 & 2.149                 & 0.6738                 \\ \hline
\multirow{2}{*}{Strategic}     & Gamer               & 1.173                 & 0.4024                 & 1.376                 & 0.4906                 & 1.862                 & 0.6235                 & 2.218                 & 0.6839                 \\ \cline{2-10} 
                               & Non-Gamer           & 1.337                 & 0.4757                 & 1.426                 & 0.5084                 & 1.856                 & 0.6223                 & 2.403                 & 0.7083                 \\ \hline
\end{tabular}
\end{table*}

Overall, in all categories of games, Strategic games are more stressful to subjects. By comparison of gamer and non-gamer, gamers are more likely to adopt strategic games. In the case of puzzle-type games, non-gamer people are more stressed than the gamer subject group.
The difference is also more significant here.
As gamer group are more used to solve the puzzle patterns it is less stressful to them. Same goes for strategic game also. But in the case of strategic game the action / shooting scenes of games are almost have same impact on both type of subjects.
The difference is also been observed in the topographic plot of brain activity. In the topographic plot, the blue region indicates the dominance of alpha activity and the red region denotes the dominance of beta activity. After only 15 minutes of gameplay, the frontal lobe of non-gamers is indicating the dominance of beta band activity. This frontal lobe is directly connected with concentration, planning, and motor control. So, this topographic plot clearly indicates that non-gamers are more impacted with puzzle gameplay. 

\begin{figure}
 \centering
\includegraphics[width=1\linewidth]{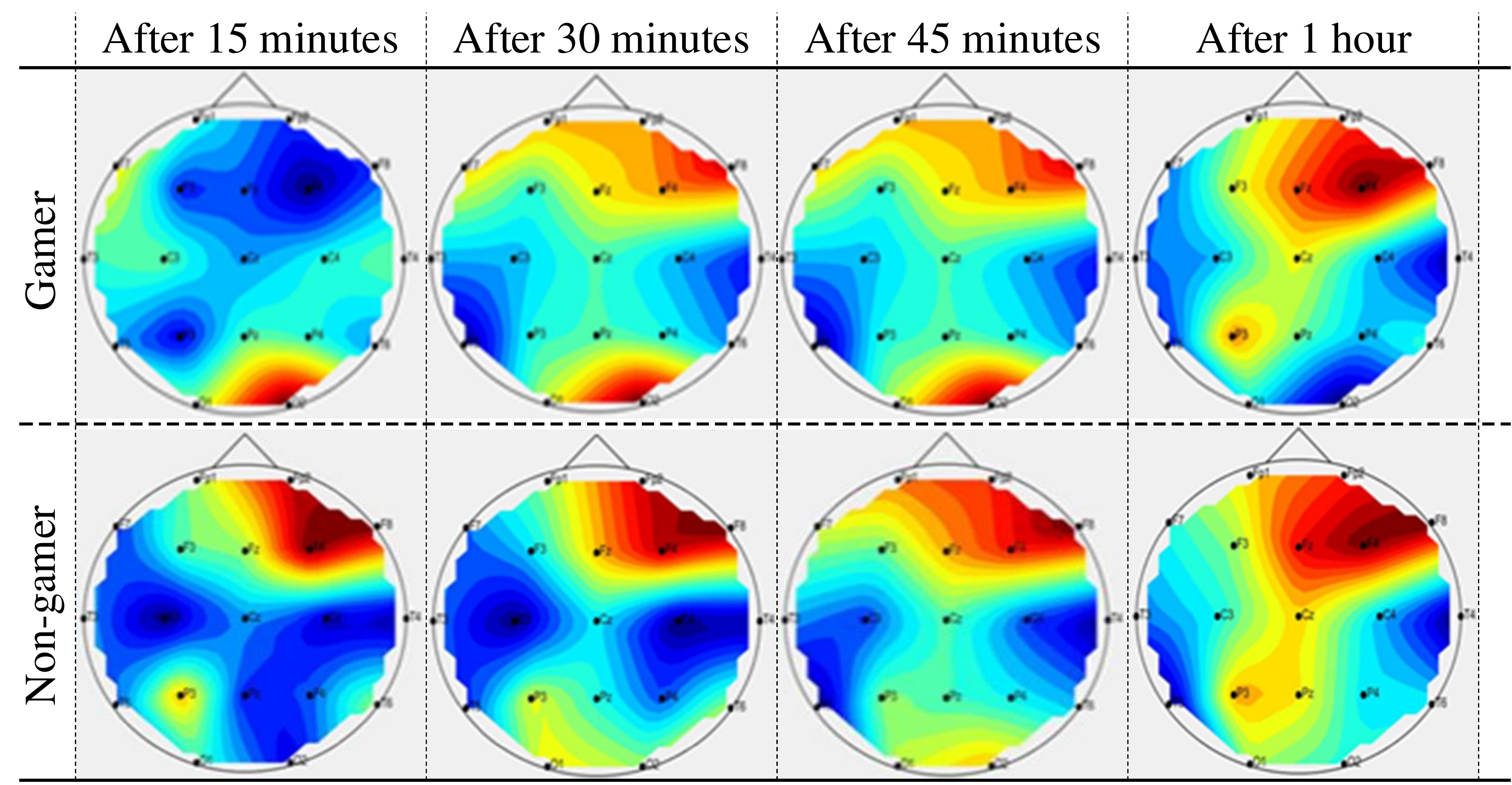}
  \caption{Topographic representation during gaming for strategic type gameplay. The blue color palette denotes the dominance of alpha activity in that region.}
  \label{fig:Topographic_representation_during}
\end{figure}

Over the gameplay time, this trend is continued to grow. Also in the same time, there is a medium hotspot of beta activity in the occipital lobe which is responsible for vision. This hotspot indicates the pressure increase in eyes for constant gameplay. And, also this region keeps growing over time. Another interesting phenomenon happens after 45 minutes of gameplay. The parietal lobe, responsible for body awareness, exhibits a mild hotspot connecting the frontal and occipital lobe. All of these brain activities clearly indicate that non-gamers show a greater beta activity over gamers. As strategic type games create the maximum stress on subjects, so in case of topographic representation only subject reaction on strategic game is depicted in Fig. \ref{fig:Topographic_representation_during}.

\subsection{Music evoked relaxation analysis}

\subsubsection{Puzzle}
In puzzle-type gameplay, the after-gameplay BAR was 1.541 for gamers and 1.897 for non-gamers.
Among the four subject groups, the group exposed to low pitch music for relaxation was more evident to bring down the beta band activity faster than the other three groups. Within this group, the beta activity in the frontal lobe and motor cortex was downed significantly faster than the other three groups. Another interesting phenomenon happened in the group, which is not exposed to no music. Their response to the brain relaxation process was just below the medium pitch song-exposed group. It is evident from the brain response that low pitch music relaxes the stressed brain faster than high pitch music. Also, at the same time, high music relaxes the brain at a much slower rate, and the residual beta-band activity is highest among all three types of music. Another noticeable change in brain activity is the creation of a mild hotspot in the temporal lobe. The change of mental state for puzzle-type gameplay is shown in Table \ref{tab:Puzzle}. The residual stress is evident in the frontal lobe when the subjects experienced high pitch music.

\begin{table*}[h]
\centering
\caption{Change of mental states after gameplay (puzzle), observed with BAR.} 
\label{tab:Puzzle}
\begin{tabular}{|m{1.5cm}|m{1.45cm}|m{2.5cm}|m{2cm}|m{2cm}|m{2cm}|m{2cm}|}
\hline
\textbf{Music Types}          & \textbf{Gamer type} & \textbf{Immediately After} & \textbf{After 3 min} & \textbf{After 6 min} & \textbf{After 9 min} & \textbf{After 12 min} \\ \hline
\multirow{2}{1.5cm}{Low pitch}    & Gamer               & 1.541                      & 1.023                & 0.865                & 0.812                & 0.709                 \\ \cline{2-7} 
                              & Non-Gamer           & 1.897                      & 1.137                & 0.912                & 0.856                & 0.716                 \\ \hline
\multirow{2}{1.5cm}{Medium Pitch} & Gamer               & 1.541                      & 1.148                & 0.892                & 0.841                & 0.714                 \\ \cline{2-7} 
                              & Non-Gamer           & 1.897                      & 1.239                & 1.026                & 0.913                & 0.768                 \\ \hline
\multirow{2}{1.5cm}{High pitch}   & Gamer               & 1.541                      & 1.267                & 0.914                & 0.862                & 0.736                 \\ \cline{2-7} 
                              & Non-Gamer           & 1.897                      & 1.342                & 1.103                & 0.924                & 0.784                 \\ \hline
\multirow{2}{1.5cm}{No music}     & Gamer               & 1.541                      & 1.186                & 0.895                & 0.852                & 0.718                 \\ \cline{2-7} 
                              & Non-Gamer           & 1.897                      & 1.289                & 1.078                & 0.917                & 0.796                 \\ \hline
\end{tabular}
\end{table*}
\subsubsection{Combinational} For combinational game types, the residual stress after 12 minutes of rest was more than in puzzle-type gameplay. Table \ref{tab:Combinational} documented the mental state change for relaxation after combinational type gameplay. The relaxation curve after combinational type gameplay is between the responses after puzzle and strategic type gameplay. The mild hotspots of the frontal lobe are smaller than strategic type gameplay (Fig. \ref{fig:Topographic_representation_after}). Another notable observation is when subjects were not experienced with any music. In this case, the response is almost similar to the medium pitch music response.

\begin{table*}[h]
\centering
\caption{Change of mental states after gameplay (combinational), observed with BAR.} 
\label{tab:Combinational}
\begin{tabular}{|m{1.5cm}|m{1.45cm}|m{2.5cm}|m{2cm}|m{2cm}|m{2cm}|m{2cm}|}
\hline
\textbf{Music Types}          & \textbf{Gamer type} & \textbf{Immediately After} & \textbf{After 3 min} & \textbf{After 6 min} & \textbf{After 9 min} & \textbf{After 12 min} \\ \hline
\multirow{2}{1.5cm}{Low pitch}    & Gamer               & 1.797                      & 1.434                & 0.968                & 0.873                & 0.713                 \\ \cline{2-7} 
                              & Non-Gamer           & 2.149                      & 1.614                & 1.172                & 0.866                & 0.72                  \\ \hline
\multirow{2}{1.5cm}{Medium Pitch} & Gamer               & 1.797                      & 1.349                & 0.98                 & 0.851                & 0.718                 \\ \cline{2-7} 
                              & Non-Gamer           & 2.149                      & 1.614                & 1.172                & 0.924                & 0.829                 \\ \hline
\multirow{2}{1.5cm}{High pitch}   & Gamer               & 1.797                      & 1.542                & 1.034                & 0.872                & 0.820                  \\ \cline{2-7} 
                              & Non-Gamer           & 2.149                      & 1.824                & 1.197                & 0.935                & 0.805                 \\ \hline
\multirow{2}{1.5cm}{No music}     & Gamer               & 1.797                      & 1.479                & 0.971                & 0.807                & 0.724                 \\ \cline{2-7} 
                              & Non-Gamer           & 2.149                      & 1.672                & 1.094                & 0.894                & 0.815                 \\ \hline
\end{tabular}
\end{table*}
\subsubsection{Strategic} 

\begin{figure}
 \centering
\includegraphics[width=1\linewidth]{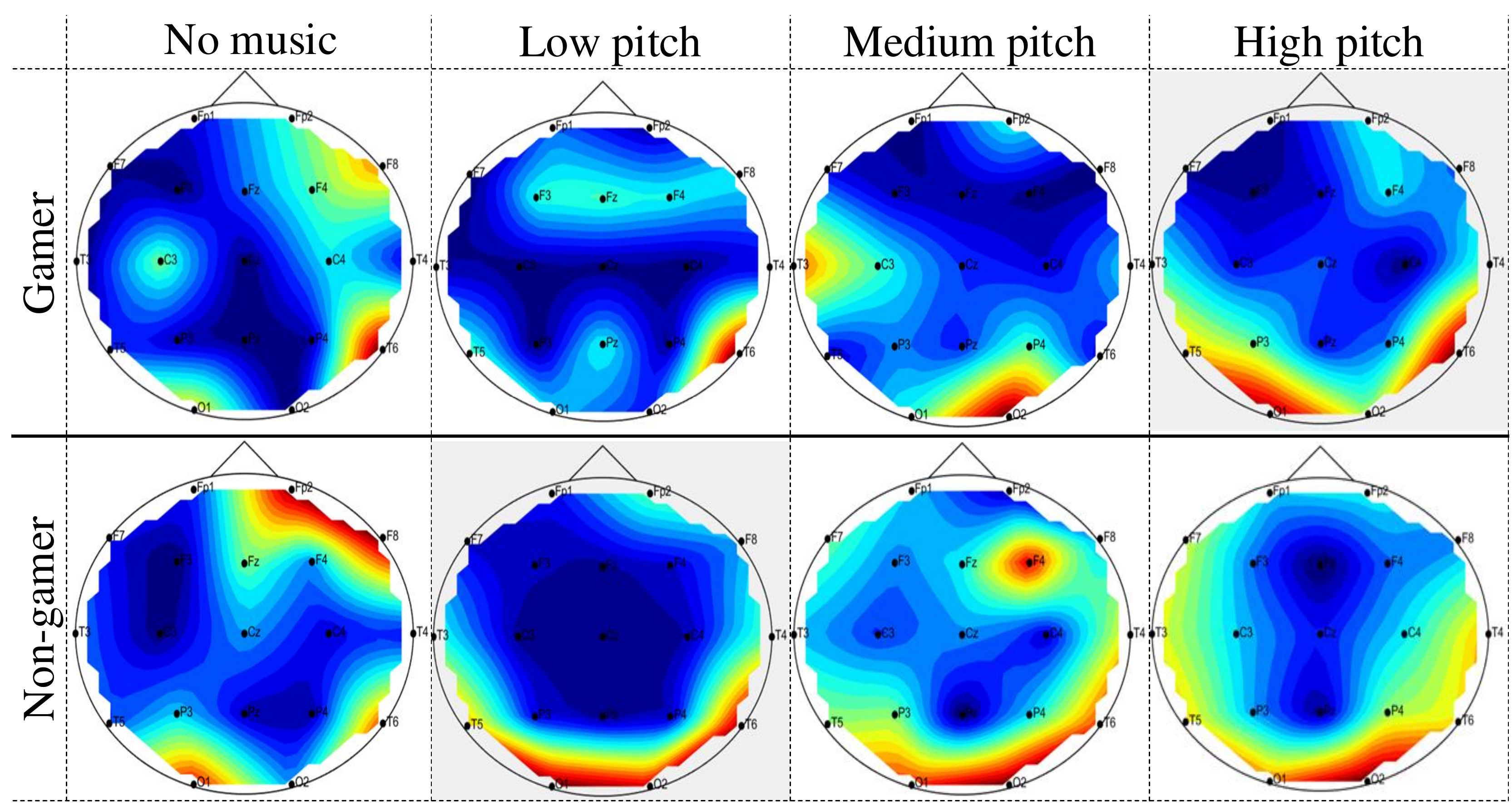}
  \caption{Topographic representation after gaming for strategic type gameplay. The red color palette denotes the dominance of beta activity in that region.}
  \label{fig:Topographic_representation_after}
\end{figure}

For strategic type gameplay, low pitch music was helpful for both gamers and non-gamers. Another interesting observation is that for non-gamers, high pitch music and no music yields an almost similar response in the relaxation curve. In Fig. \ref{fig:Topographic_representation_after} the phases of brain after 12 minutes for various cases are depicted in topographic representation. A mild hotspot of beta activity in the frontal lobe is present for non-gamers when they are experienced to medium pitch music for relaxation. Nevertheless, the overall percentage of beta activity is much higher in non-gamers who experienced high pitch music. When gamers have not experienced any music, some mild activity zones are visible frontal and occipital lobes. Overall, both for gamers and non-gamers, low pitch music brings more relaxation than any other relaxation procedure (Table \ref{tab:Strategic}).

\begin{table*}[h]
\centering
\caption{Change of mental states after gameplay (strategic), observed with BAR.} 
\label{tab:Strategic}
\begin{tabular}{|m{1.5cm}|m{1.45cm}|m{2.5cm}|m{2cm}|m{2cm}|m{2cm}|m{2cm}|}
\hline
\textbf{Music Types}          & \textbf{Gamer type} & \textbf{Immediately After} & \textbf{After 3 min} & \textbf{After 6 min} & \textbf{After 9 min} & \textbf{After 12 min} \\ \hline
\multirow{2}{1.5cm}{Low pitch}    & Gamer               & 2.218                      & 1.451                & 1.145                & 0.814                & 0.717                 \\ \cline{2-7} 
                              & Non-Gamer           & 2.403                      & 1.663                & 1.232                & 0.873                & 0.749                 \\ \hline
\multirow{2}{1.5cm}{Medium Pitch} & Gamer               & 2.218                      & 1.71                 & 1.205                & 0.917                & 0.723                 \\ \cline{2-7} 
                              & Non-Gamer           & 2.403                      & 1.846                & 1.386                & 0.996                & 0.824                 \\ \hline
\multirow{2}{1.5cm}{High pitch}   & Gamer               & 2.218                      & 1.76                 & 1.235                & 0.94                 & 0.816                 \\ \cline{2-7} 
                              & Non-Gamer           & 2.403                      & 1.999                & 1.49                 & 1.008                & 0.827                 \\ \hline
\multirow{2}{1.5cm}{No music}     & Gamer               & 2.218                      & 1.75                 & 1.296                & 0.951                & 0.741                 \\ \cline{2-7} 
                              & Non-Gamer           & 2.403                      & 1.983                & 1.472                & 1.017                & 0.819                 \\ \hline
\end{tabular}
\end{table*}

\begin{figure*}
 \centering
\includegraphics[width=1\linewidth]{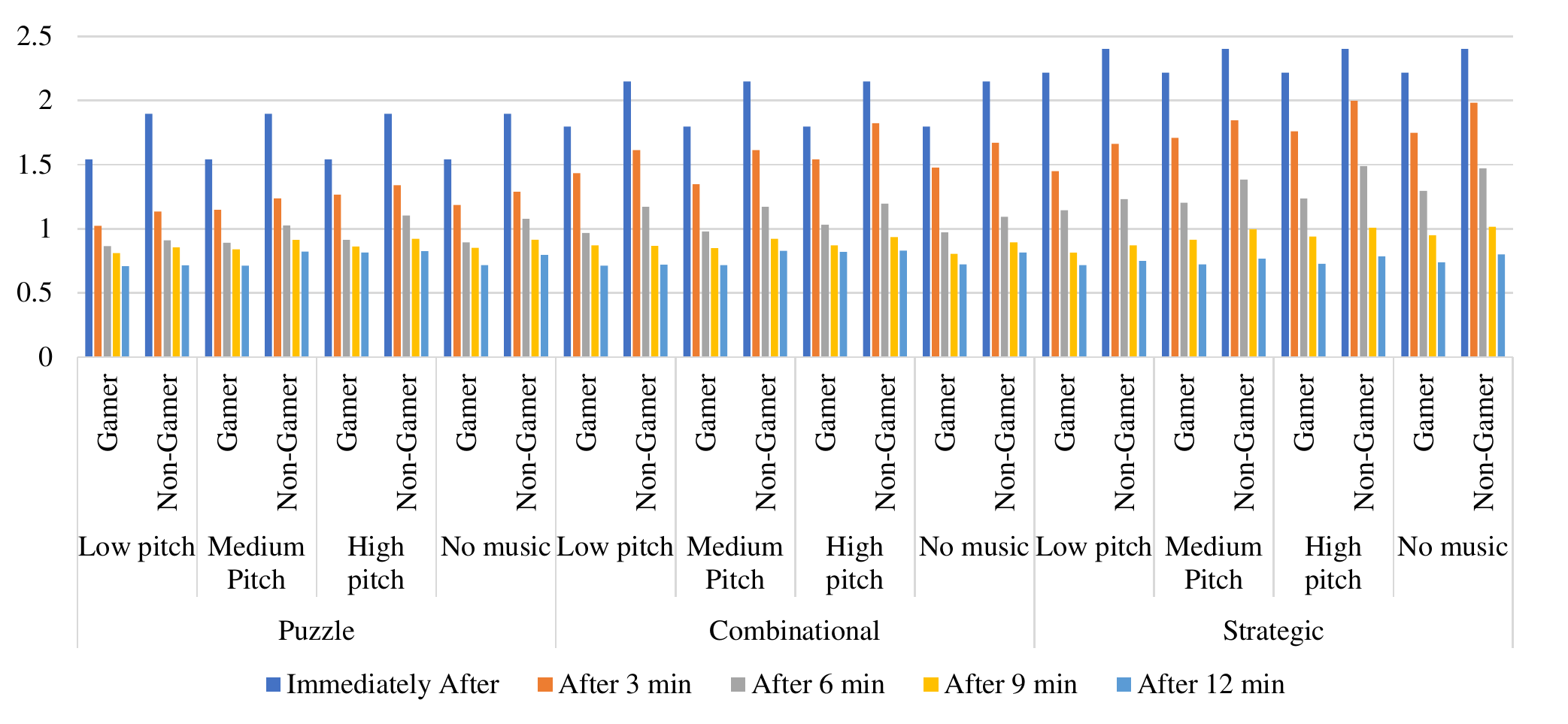}
  \caption{Brain relaxation after gameplay.}
  \label{fig:Mental_states_after_gameplay_(Strategic)}
\end{figure*} 




\subsection{Regression based mathematical analysis}
For the mathematical modelling of the entire experiment, two different phases have been followed, such as:

\begin{enumerate}
  \item \textbf{During gameplay:} Tables \ref{tab:math_during_4pl} and \ref{tab:math_during_4q} denote the parameters for 4PL symmetrical sigmoid function and quartic regression, respectively. $R^2$ and AIC is used to measure the goodness of the mathematical modeling for the respective equations. In the later gameplay period, stress increase rate was much higher than the earlier time.
  
  \item \textbf{After gameplay:} In Tables  \ref{tab:after_4PL} and \ref{tab:after_Quartic}, the mathematical modeling for brain relaxation is documented. In the first three minutes the rate of relaxation is higher than the later segments.
\end{enumerate}

\begin{figure}
 \centering
\includegraphics[width=0.8\linewidth]{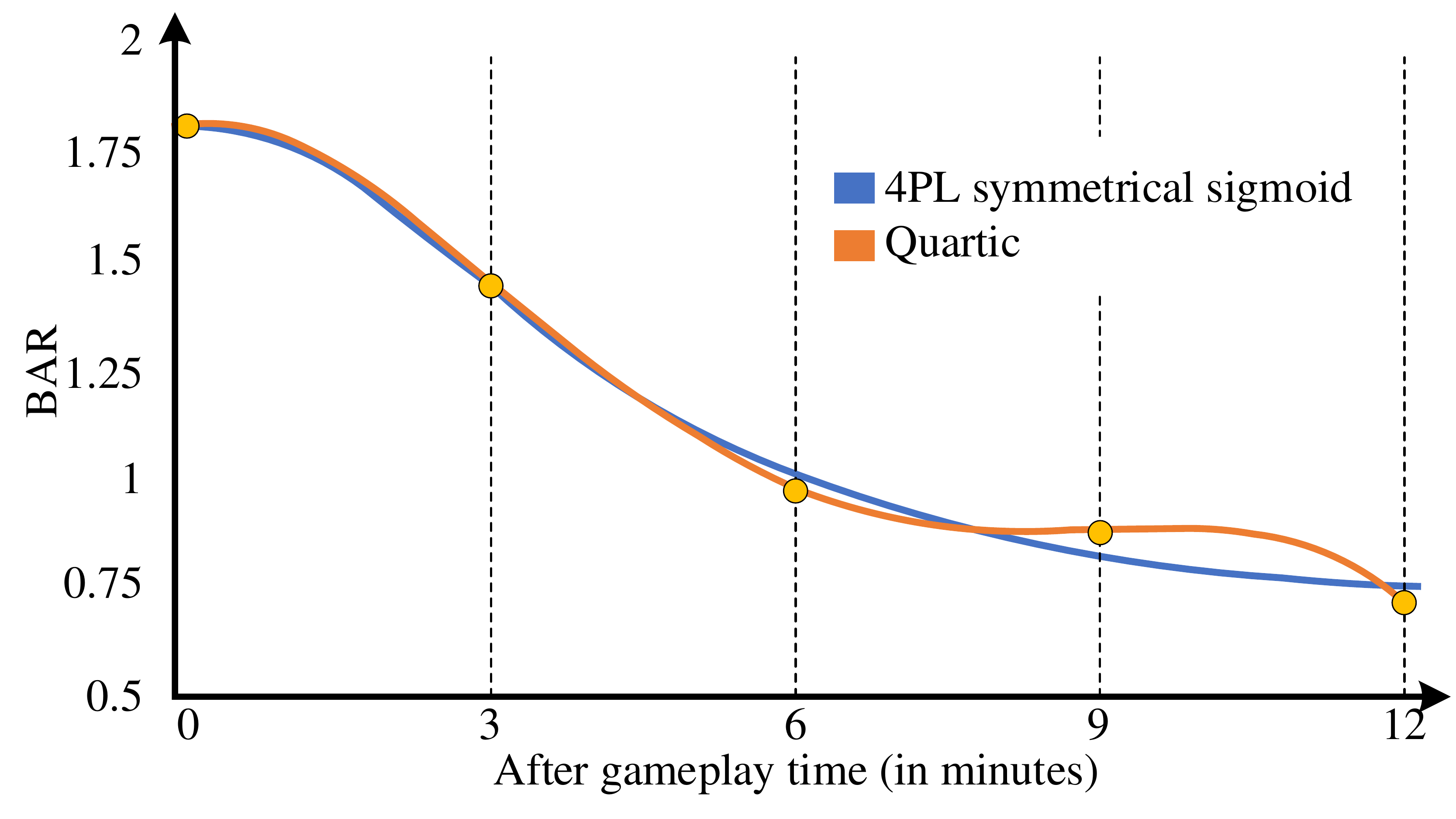}
  \caption{Residuals of cubic and original average data}
  \label{fig:eqn_comp}
\end{figure}

In Fig. \ref{fig:eqn_comp}, the comparison between the two mathematical models is depicted. In almost all cases, the quartic regression dominates over the 4PL symmetrical sigmoid function. But this achievement has come with a price for quartic regression. On average, 20-30\% more computational power was required to model quartic regression than the 4PL symmetrical sigmoid function.

\begin{table}[!ht]
\centering
\caption{Parameters and metrics of mathematical modeling for during gameplay experiment (4PL symmetrical sigmoid function).}
\label{tab:math_during_4pl}
\begin{tabular}{|m{2cm}|m{1.45cm}|m{1.65cm}|m{1.65cm}|m{1.65cm}|m{1.45cm}|m{1.45cm}|m{1.45cm}|}
\hline
\multirow{2}{2cm}{\textbf{Game type}} & \multirow{2}{1.45cm}{\textbf{Gamer type}} & \multicolumn{4}{c|}{\textbf{Coefficient}}         & \multicolumn{2}{c|}{\textbf{Metrics}} \\ \cline{3-8} 
                                     &                                      & \textit{a} & \textit{b} & \textit{c} & \textit{d} & \textit{$R^2$}       & \textit{AIC}      \\ \hline
\multirow{2}{*}{Puzzle}              & Gamer                                & 0.7113  & 5.0082  & 41.0507   & 1.6653   & 0.9996            & -27.47            \\ \cline{2-8} 
                                     & Non Gamer                            & 0.7701  & 4.5143   & 42.8531   & 2.1471   & 0.9887            & -8.195            \\ \hline
\multirow{2}{2cm}{Combi- national}       & Gamer                                & 0.7371   & 3.1548     & 62.6104   & 3.0142   & 0.9943            & -12.68            \\ \cline{2-8} 
                                     & Non Gamer                            & 0.7223  & 1.1433   & 7008    & 8794     & 0.9919            & -8.719            \\ \hline
\multirow{2}{*}{Strategic}           & Gamer                                & 0.7187  & 0.9913  & 2403   & 5339054    & 0.989             & -6.724            \\ \cline{2-8} 
                                     & Non Gamer                            & 0.7472  & 0.9616  & 290172   & 460270.4   & 0.9605            & 0.3441            \\ \hline
\end{tabular}
\end{table}

\begin{table}[!ht]
\centering
\caption{Parameters and metrics of mathematical modeling for during gameplay experiment (quartic regression).}
\label{tab:math_during_4q}
\begin{tabular}{|m{1.5cm}|m{1.5cm}|m{1cm}|m{1.7cm}|m{1.7cm}|m{1.9cm}|m{2cm}|m{1.5cm}|}
\hline
\multirow{2}{1.5cm}{\textbf{Game type}} & \multirow{2}{1.5cm}{\textbf{Gamer type}} & \multicolumn{5}{c|}{\textbf{Coefficient}}                           & \textbf{Metrics} \\ \cline{3-8} 
                                     &                                      & \textit{a} & \textit{b} & \textit{c}   & \textit{d}  & \textit{e}   & \textit{$R^2$}      \\ \hline
\multirow{2}{*}{Puzzle}              & Gamer                                & 0.701      & 0.01184 & -0.00136 & 5.373E-5  & -5.086E-7 & 1                \\ \cline{2-8} 
                                     & Non Gamer                            & 0.701      & 0.04116 & -0.00341 & 10.598E-5 & -9.169E-7 & 1                \\ \hline
\multirow{2}{2cm}{Combi- national}       & Gamer                                & 0.701      & 0.02556 & -0.00202 & 6.227E-5 & -5.103E-7 & 1                \\ \cline{2-8} 
                                     & Non Gamer                            & 0.701      & 0.05173 & -0.00268 & 7.081E-5 & -5.629E-7 & 1                \\ \hline
\multirow{2}{*}{Strategic}           & Gamer                                & 0.701      & 0.06878 & -0.00379 & 9.874E-5 & -7.942E-7 & 1                \\ \cline{2-8} 
                                     & Non Gamer                            & 0.701      & 0.0989     & -54541     & 0.000126 & -9.152E-7 & 1                \\ \hline
\end{tabular}
\end{table}

\begin{table*}[!ht]
\tiny
\caption{After gameplay mathematical modeling with 4PL symmetrical sigmoid function.}
\label{tab:after_4PL}
\begin{tabular}{|m{1.5cm}|m{1.45cm}|m{1.45cm}|m{1.45cm}|m{1.45cm}|m{1.45cm}|m{1.45cm}|m{1.45cm}|m{1.45cm}|}
\hline
\multirow{2}{1.5cm}{\textbf{Game type}} & \multirow{2}{1.45cm}{\textbf{Music type}} & \multirow{2}{1.45cm}{\textbf{Gamer type}} & \multicolumn{4}{l|}{\textbf{Coefficient}}          & \multicolumn{2}{l|}{\textbf{Metrics}} \\ \cline{4-9} 
                                    &                                      &                                      & \textit{a} & \textit{b} & \textit{c} & \textit{d}  & \textit{$R^2$}       & \textit{AIC}      \\ \hline
\multirow{8}{*}{Puzzle}             & \multirow{2}{1.45cm}{Low Pitch}           & Gamer                                & 1.5410   & 0.3937  & 360.9098   & -2.4249   & 0.9979            & -20.95            \\ \cline{3-9} 
                                    &                                      & Non Gamer                            & 1.8970   & 0.4967 & 14.7121   & -0.5570   & 0.9973            & -16.05            \\ \cline{2-9} 
                                    & \multirow{2}{1.45cm}{Medium Pitch}        & Gamer                                & 1.5414   & 1.2729   & 4.2799   & 0.5133   & 0.9937            & -15.37            \\ \cline{3-9} 
                                    &                                      & Non Gamer                            & 1.8970   & 0.79044  & 5.1763   & 0.2516   & 0.9999            & -36.46            \\ \cline{2-9} 
                                    & \multirow{2}{1.45cm}{High Pitch}          & Gamer                                & 1.5412   & 3.2904  & 3.4865   & 0.8151    & 0.9989            & -24.73            \\ \cline{3-9} 
                                    &                                      & Non Gamer                            & 1.8969   & 0.8284   & 8.9388    & -0.0215 & 0.0009            & -28.03            \\ \cline{2-9} 
                                    & \multirow{2}{1.45cm}{No music}            & Gamer                                & 1.5417   & 1.6380    & 3.9345   & 0.6141   & 0.9913            & -13.77            \\ \cline{3-9} 
                                    &                                      & Non Gamer                            & 1.8969   & 0.4707   & 922.3299   & -7.7168   & 1                 & -40.25            \\ \hline
\multirow{8}{1.5cm}{Combi- national}      & \multirow{2}{1.45cm}{Low Pitch}           & Gamer                                & 1.7987   & 2.2673   & 4.1514  & 0.6491   & 0.994             & -12.53            \\ \cline{3-9} 
                                    &                                      & Non Gamer                            & 2.1484   & 1.4688   & 5.8055   & 0.2179   & 0.9996            & -23.86            \\ \cline{2-9} 
                                    & \multirow{2}{1.45cm}{Medium Pitch}        & Gamer                                & 1.7975    & 1.6254   & 4.3045   & 0.5303   & 0.9984            & -19.37            \\ \cline{3-9} 
                                    &                                      & Non Gamer                            & 2.1486   & 1.7268   & 4.3938   & 0.5875   & 0.9997            & -25.68            \\ \cline{2-9} 
                                    & \multirow{2}{1.45cm}{High Pitch}          & Gamer                                & 1.7970   & 2.3198    & 3.3399    & 0.7732   & 0.9999            & -35.66            \\ \cline{3-9} 
                                    &                                      & Non Gamer                            & 2.1487   & 1.7094   & 4.4000    & 0.5887   & 0.9999            & -31               \\ \cline{2-9} 
                                    & \multirow{2}{1.45cm}{No music}            & Gamer                                & 1.7975   & 2.7664   & 4.1841   & 0.6739   & 0.9997            & -27.28            \\ \cline{3-9} 
                                    &                                      & Non Gamer                            & 2.1490   & 2.5295   & 3.9101   & 0.7385   & 1                 & -42.42            \\ \hline
\multirow{8}{*}{Strategic}          & \multirow{2}{1.45cm}{Low Pitch}           & Gamer                                & 2.2175   & 0.8197  & 0.89861    & -0.7128  & 0.9965            & -12.21            \\ \cline{3-9} 
                                    &                                      & Non Gamer                            & 2.4024    & 1.1049   & 7.2473   & -0.2689  & 0.9988            & -16.5             \\ \cline{2-9} 
                                    & \multirow{2}{1.45cm}{Medium Pitch}        & Gamer                                & 2.2183   & 1.5853   & 5.8047   & 0.2552   & 0.9999            & -30.82            \\ \cline{3-9} 
                                    &                                      & Non Gamer                            & 2.4022   & 1.2444   & 9.5813   & -0.4817  & 0.9995            & -21.38            \\ \cline{2-9} 
                                    & \multirow{2}{1.45cm}{High Pitch}          & Gamer                                & 2.2188   & 1.6827   & 5.9997   & 0.2723595   & 0.9998            & -25.39            \\ \cline{3-9} 
                                    &                                      & Non Gamer                            & 2.399616   & 1.754762   & 7.592323   & 0.03812683  & 0.9985            & -15.49            \\ \cline{2-9} 
                                    & \multirow{2}{1.45cm}{No music}            & Gamer                                & 2.217533   & 1.426625   & 7.804479   & -0.06470811 & 0.9999            & -28.79            \\ \cline{3-9} 
                                    &                                      & Non Gamer                            & 2.400454   & 1.721276   & 7.344441   & 0.09479044  & 0.9989            & -17.12            \\ \hline
\end{tabular}
\end{table*}

\begin{table*}[!ht]
\tiny
\caption{After gameplay mathematical modeling with Quartic Regression function.}
\label{tab:after_Quartic}
\begin{tabular}{|m{1.45cm}|m{1.45cm}|m{1.45cm}|m{1.2cm}|m{1.45cm}|m{1.45cm}|m{1.45cm}|m{1.6cm}|m{1.45cm}|}
\hline
\multirow{2}{1.45cm}{\textbf{Game type}} & \multirow{2}{1.45cm}{\textbf{Music type}} & \multirow{2}{1.45cm}{\textbf{Gamer type}} & \multicolumn{5}{c|}{\textbf{Coefficient}}                             & \textbf{Metrics} \\ \cline{4-9} 
                                    &                                      &                                      & \textit{a} & \textit{b}  & \textit{c}   & \textit{d}   & \textit{e}   & \textit{$R^2$}               \\ \hline
\multirow{8}{*}{Puzzle}             & \multirow{2}{1.45cm}{Low Pitch}           & Gamer                                & 1.541      & -0.2693  & 0.0393   & -0.0025      & 0.000051     & 1                \\ \cline{3-9} 
                                    &                                      & Non Gamer                            & 1897       & -0.3926  & 0.0558   & -0.00331 & 5.81276E-5  & 1                \\ \cline{2-9} 
                                    & \multirow{2}{1.45cm}{Medium Pitch}        & Gamer                                & 1.541      & -0.1170  & -0.01404  & 0.00367  & -0.00018 & 1                \\ \cline{3-9} 
                                    &                                      & Non Gamer                            & 1.897      & -0.3543    & 0.0576   & -0.0046  & 0.00014  & 1                \\ \cline{2-9} 
                                    & \multirow{2}{1.45cm}{High Pitch}          & Gamer                                & 1.541      & 0.0203  & -0.0599    & 0.0086  & -0.00035 & 1                \\ \cline{3-9} 
                                    &                                      & Non Gamer                            & 1.897      & -0.2892  & 0.0459   & -0.00415 & 0.00014   & 1                \\ \cline{2-9} 
                                    & \multirow{2}{1.45cm}{No music}            & Gamer                                & 1.541      & -0.0649 & -0.0333  & 0.00598  & -0.00027 & 1                \\ \cline{3-9} 
                                    &                                      & Non Gamer                            & 1.897      & -0.3355  & 0.0585   & -0.0053 & 0.00017  & 1                \\ \hline
\multirow{8}{1.45cm}{Combi- national}      & \multirow{2}{1.45cm}{Low Pitch}           & Gamer                                & 1.797      & 0.0247   & -0.0784  & 0.0114   & -0.00047 & 1                \\ \cline{3-9} 
                                    &                                      & Non Gamer                            & 2.149      & -0.1875  & 0.00181  & 0.00044  & -9.774E-06 & 1                \\ \cline{2-9} 
                                    & \multirow{2}{1.45cm}{Medium Pitch}        & Gamer                                & 1.797      & -0.1109  & -0.02518  & 0.00474  & -0.00021 & 1                \\ \cline{3-9} 
                                    &                                      & Non Gamer                            & 2.149      & -0.1708  & -0.0077 & 0.00194  & -7.305E-5 & 1                \\ \cline{2-9} 
                                    & \multirow{2}{1.45cm}{High Pitch}          & Gamer                                & 1.797      & -0.11003  & -0.0254  & 0.0047   & -0.00019 & 1                \\ \cline{3-9} 
                                    &                                      & Non Gamer                            & 2.149      & -0.1661  & -0.0104  & 0.0024  & -9.465E-5 & 1                \\ \cline{2-9} 
                                    & \multirow{2}{1.45cm}{No music}            & Gamer                                & 1.797      & 0.0514 & -0.0808  & 0.0107   & -0.00041 & 1                \\ \cline{3-9} 
                                    &                                      & Non Gamer                            & 2.149      & -0.0276 & -0.069   & 0.0098  & -0.00038 & 1                \\ \hline
\multirow{8}{*}{Strategic}          & \multirow{2}{1.45cm}{Low Pitch}           & Gamer                                & 2.218      & -0.4486  & 0.0906   & -0.00989 & 0.00038  & 1                \\ \cline{3-9} 
                                    &                                      & Non Gamer                            & 2.403      & -0.3553  & 0.04918   & 0.00019  & -0.00098 & 1                \\ \cline{2-9} 
                                    & \multirow{2}{1.45cm}{Medium Pitch}        & Gamer                                & 2.218      & -0.1179  & -0.0289  & 0.0044  & -0.00017 & 1                \\ \cline{3-9} 
                                    &                                      & Non Gamer                            & 2.403      & -0.21475    & 0.0129   & -0.00127 & 6.121E-5   & 1                \\ \cline{2-9} 
                                    & \multirow{2}{1.45cm}{High Pitch}          & Gamer                                & 2.218      & -0.0715     & -0.0428  & 0.0059  & -0.00023 & 1                \\ \cline{3-9} 
                                    &                                      & Non Gamer                            & 2.403      & -0.11075    & -0.008    & -0.0001019 & 5.0926E-5  & 1                \\ \cline{2-9} 
                                    & \multirow{2}{1.45cm}{No music}            & Gamer                                & 2.218      & -0.1420278  & -0.00801 & 0.00123  & -3.549E-5 & 1                \\ \cline{3-9} 
                                    &                                      & Non Gamer                            & 2.403      & -0.1115     & -0.01139  & 0.0006  & 1.852E-5  & 1                \\ \hline
\end{tabular}
\end{table*}

The strategic game creates more stress on the subjects of the three broad categories of games discussed here. However, gamer subject groups adapted more to strategic games because of their experience.

In the relaxation phase, the subject groups who experienced puzzle games achieved less BAR than other groups. Even for non-gamers, this observation stands out. We conclude that even when subjects were not experienced to puzzle games, puzzles are commonplace in our lives. So, the brain seems to adapt too quickly.

For mathematical modeling, during gameplay, the stress curve is very predictable, resulting in $R^2$ value 1 for every case. However, when 4PL symmetrical sigmoid functions were used to model relaxation curves, the $R^2$ value drops to even 0.9913. Only in two cases, the  $R^2$ value attains 1. In quartic regression function, despite requiring much more computational power, yields the  $R^2$ value to 1.

\section{Outcome of the Study}
The key outcome of the study is delineated as follows:

\begin{enumerate}
  \item Non-gamers experience more stress than gamers,
  
  \item During gameplay, the beta band in the frontal region is mostly activated,
  
  \item	Strategic game creates more stress on human brain,
  
  \item Low pitch music is the most useful media for relaxation,
  
  \item Residual stress is evident in the frontal lobe when the subjects experienced high pitch music,
  
  \item During relaxation, the alpha band in the frontal region is mostly activated,
  
  \item 4PL symmetrical sigmoid function performs regression analysis with minimum parameters/ computational power,
  
  \item	Quartic regression performs regression analysis follow the relaxation curve more accurately compared to 4PL symmetrical sigmoid function.
\end{enumerate}

\section{Conclusion}
The study of stress analysis on the human brain for video gaming and music-evoked relaxation analysis is critical for understanding the human brain for everyday activities. This study is, to the best of our knowledge, the first of its kind. This paper presents a comprehensive stress analysis for video gameplay from a wide range of perspectives. Stress was analyzed from the rhythm-specific spectral feature, BAR, from EEG. Among all games, the strategic game proves to be the most stressful. Non-gamers exhibit more stress to video gameplay than gamers, demonstrating the adaptability of video gameplay with stress. Music of various pitches was used to compare no music after the gameplay for relaxation analysis. After video gameplay, low pitch music proved to be the most effective media for relaxation. Regression analysis was used to represent relaxation and stress mathematically better. The most effective function for this is quartic regression. EEG topography has been used to identify the activation hotspot within the brain for a specific game or relaxation procedure. During gameplay, the frontal region is particularly active. This work can be expanded to include other types of VG or even other stressful activities. In addition, any relaxation technique other than music can be used to normalize brain activity.

\section*{Acknowledgements}
None. No funding to declare.

\section*{Declaration of Competing Interest}
The authors have no conflict of interest to disclose.

\bibliographystyle{model2-names}

\bibliography{sample}

\end{document}